\documentclass[pra,twocolumn,showpacs,floatfix]{revtex4-1}
\usepackage{amsmath}
\usepackage{epsfig}
\usepackage{amssymb}
\usepackage{dcolumn}
\usepackage{graphicx}
\usepackage{dcolumn}

\begin{document}

\date{\today }
\title{Localization and delocalization of two-dimensional discrete solitons
pinned to linear and nonlinear defects}
\author{Valeriy A. Brazhnyi}
\email{brazhnyy@gmail.com}
\affiliation{Centro de F\'{\i}sica do Porto, Faculdade de Ci\^encias, Universidade do
Porto, R. Campo Alegre 687, Porto 4169-007, Portugal}
\author{Boris A. Malomed}
\email{malomed@post.tau.ac.il}
\affiliation{Department of Physical Electronics, Faculty of Engineering, Tel Aviv
University, Tel Aviv 69978, Israel}

\begin{abstract}
We study the dynamics of two-dimensional (2D) localized modes in the
nonlinear lattice described by the discrete nonlinear Schr\"{o}dinger (DNLS)
equation, including a local linear or nonlinear defect. Discrete solitons
pinned to the defects are investigated by means of the numerical
continuation from the anti-continuum limit and also using the variational
approximation (VA), which features a good agreement for strongly localized
modes. The models with the time-modulated strengths of the linear or
nonlinear defect are considered too. In that case, one can temporarily shift
the critical norm, below which localized 2D modes cannot exists, to a level
above the norm of the given soliton, which triggers the irreversible
delocalization transition.
\end{abstract}

\pacs{03.75.Lm, 03.75.Kk, 03.75.-b}
\maketitle

\section{Introduction}

The discrete nonlinear Schr\"{o}dinger (DNLS)\ equations constitute a
universal class of models which are profoundly interesting in their own
right, as dynamical systems, and serve as models of a number of physical
systems in nonlinear optics \cite{PhysRep}, studies of Bose-Einstein
condensates \cite{Smerzi, morsch}, and in other contexts \cite{Panos}. In
particular, soliton solutions to the DNLS equation in one, two, and three
dimensions represent fundamental dynamically localized modes in discrete
media. Experimentally, one- and two-dimensional (1D and 2D) discrete
solitons have been created in nonlinear optical systems of several types
\cite{PhysRep}.

An important ingredient of DNLS models is represented by local defects. They
are interesting as additional elements of the lattices \cite{Panos}, and
find important physical realizations. In particular, they may describe
various strongly localized structures in photonic crystals \cite{defect0},
including nanocavities \cite{defect1}, as well as micro-resonators \cite%
{defect2}, and quantum dots \cite{defect3}.

The objective of the present work is to consider the interaction of 2D
discrete solitons with local linear and nonlinear defects in DNLS lattices.
The defect may be concentrated at a single site of the lattice, or it may be
shaped as a Gaussian of a finite width. After introducing the model in
Section 2, we consider stationary solitons pinned by the defects in Section
3. The analysis is performed by means of the variational approximation (VA)
and numerical methods. In particular, the pinned-soliton families feature a
specific combination of stable and unstable parts. In Section 4, we consider
possibilities of the application of the ``management" \cite%
{book} to 2D discrete solitons, using the linear and nonlinear defects whose
amplitudes slowly vary in time. In this direction, we analyze a possibility
to trigger a delocalization transition by means of this method, which may be
used in the design of switching schemes in photonics. Previously, the
induced transition to the delocalization was demonstrates in uniform
lattices with the inter-site coupling strength subject to the time
modulation \cite{Bishop}. The paper is concluded by Section 5.

\section{The model}

We consider the following model based on the 2D DNLS equation with a local
defect:
\begin{eqnarray}
i\dot{u}_{n,m} &+&J\Delta _{2}u_{n,m}+V_{n,m}u_{n,m}  \notag \\
&+&(\sigma +W_{n,m})|u_{n,m}|^{2}u_{n,m}=0,  \label{timeDNLS}
\end{eqnarray}%
where the overdot stands for the time derivative, $\Delta _{2}u_{n,m}\equiv
u_{n,m+1}+u_{n,m-1}+u_{n+1,m}+u_{n-1,m}-4u_{n,m}$ is the 2D discrete
Laplacian, the coupling constant of the lattice will be fixed by scaling, $%
J\equiv 1$, unless another choice of $J$ is specified explicitly, and $%
\sigma =+1$ and $-1$ corresponds to the attractive and repulsive
nonlinearities, respectively ($\sigma =0$ for the linear lattice). Further,
functions $V_{n,m}$ and $W_{n,m}$, which account for the linear and
nonlinear components of the defect, are taken as Gaussians profiles,
\begin{eqnarray}
V_{n,m} &=&v\exp \left[ -(n^{2}+m^{2})/\Delta _{v}\right] ,  \notag \\
W_{n,m} &=&w\exp \left[ -(n^{2}+m^{2})/\Delta _{w}\right] ,  \label{defects}
\end{eqnarray}%
with respective strengths $v,w$ and widths $\Delta _{v},\Delta _{w}$. In
this notation, positive and negative strengths correspond to the attractive
and repulsive defect, respectively. In fact, we will consider the linear and
nonlinear defects separately. Models of 2D nonlinear lattices with other
types of local defects were considered earlier \cite{Gaididei}, including
defects induced by edges of the lattice \cite{Kivshar}.

Looking for the stationary solutions,
\begin{equation}
u_{n,m}=U_{n,m}\exp (-i\omega t),  \label{omega}
\end{equation}%
we arrive at the nonlinear eigenvalue problem,
\begin{eqnarray}
\omega U_{n,m} &+&\Delta _{2}U_{n,m}+V_{n,m}U_{n,m}  \notag \\
&+&(\sigma +W_{n,m})|U_{n,m}|^{2}U_{n,m}=0,  \label{steadyDNLS}
\end{eqnarray}%
for real frequency $\omega $ and the profile of the stationary discrete mode
$U_{m,n}$, which may be complex, in the general case. In the absence of the
defect, Eq. (\ref{steadyDNLS}) gives rise to the linear dispersion relation
featuring the phonon band, above and below which one can find nonlinear
modes, described by respective curves $N(\omega )$, with $%
N=\sum_{m,n}|u_{m,n}|^{2}$ being the norm (power) of the nonlinear state.
The important difference between 2D and 1D settings is that, in the latter
case, the fundamental single-peak mode (the discrete soliton of the
Sievers-Takeno type) persists in the limit of $N\rightarrow 0$, while all
the 2D solitons are bounded by a critical norm, $N_{\mathrm{cr}}$, below
which localized modes do not exist \cite{Panos}. Accordingly, an abrupt
delocalization (decay) of discrete 2D solitons was predicted in the case
when the inter-site coupling constant would exceed a certain critical value
\cite{Bishop}. In the following we show that, introducing the defect with
the linear and nonlinear components and varying their strengths, or the
lattice coupling constant, one can change the critical norm, $N_{\mathrm{cr}%
} $, and thus control the transition to the delocalization.

\section{Stationary discrete solitons pinned to the defect: The variational
approach and numerical results}

\subsection{The variational approximation}

The variational approach (VA) was successfully used for the study of 2D
discrete solitons in uniform (defect-free) DNLS lattices \cite{Weinstein,CQ}%
. Here, we start with the application of the VA to the 2D lattice in the
presence of the $\delta$-defect localized at the origin:
\begin{gather}
i\dot{u}_{m,n}+\Delta _{2}u_{m,n}+\sigma |u_{m,n}|^{2}u_{m,n}  \notag \\
+(v+w|u_{m,n}|^{2})\delta _{m,0}\delta _{n,0}u_{m,n}=0,  \label{eq:dnls_def}
\end{gather}%
cf. the more general form of the defect in Eq. (\ref{defects}). Generally,
the strength of the defect may be time-dependent, $v=v(t),~w=w(t)$.

Equation (\ref{eq:dnls_def}) can be derived from the Lagrangian,%
\begin{eqnarray}
L &=&\frac{i}{2}\sum_{m,n}\left( u_{m,n}^{\ast }\dot{u}_{m,n}-u_{m,n}\dot{u}%
_{m,n}^{\ast }\right)  \notag \\
&+&\sum_{m,n}\left[ u_{m,n}^{\ast }\left( u_{m+1,n}+u_{m,n+1}\right) \right.
\notag \\
&+&\left. u_{m,n}\left( u_{m+1,n}^{\ast }+u_{m,n+1}^{\ast }\right)
-4|u_{m,n}|^{2}\right]  \notag \\
&+&\frac{\sigma }{2}\sum_{m,n}|u_{m,n}|^{4}+v|u_{0,0}|^{2}+\frac{1}{2}%
w|u_{0,0}|^{4}.  \label{eq:lagrag}
\end{eqnarray}%
To apply the VA, we adopt the following ansatz (previously, it was used for
the analysis of discrete solitons in the 2D DNLS equation with the
cubic-quintic nonlinearity \cite{CQ}):
\begin{equation}
u_{m,n}=\left\{
\begin{array}{ll}
B, & \mbox{if}\quad m=n=0; \\
Ae^{-a(|m|+|n|)}, & \mbox{otherwise}.%
\end{array}%
\right.  \label{eq:ansatz}
\end{equation}

Substituting ansatz (\ref{eq:ansatz}) into Lagrangian (\ref{eq:lagrag}), one
can perform the respective calculations and find the effective Lagrangian as
a sum of four terms which depend on three dynamical parameters, $(A,B,a)$
(which may be functions of time) and represent, respectively, the kinetic
part, on-site self-interaction, inter-site couplings, and effects induced by
the defect:%
\begin{gather}
L=L_{\mathrm{kin}}+L_{\mathrm{coupl}}+L_{\mathrm{int}}+L_{\mathrm{self}}~; \\
L_{\mathrm{kin}}=\frac{2ie^{-2a}}{\left( 1-e^{-2a}\right) ^{2}}\left(
A^{\ast }\dot{A}-A\dot{A}^{\ast }\right) +\frac{i}{2}\left( B^{\ast }\dot{B}%
-B\dot{B}^{\ast }\right) , \\
L_{\mathrm{self}}=2\sigma |A|^{4}\frac{e^{-4a}}{\left( 1-e^{-4a}\right) ^{2}}%
+\frac{\sigma }{2}|B|^{4}, \\
L_{\mathrm{coupl}}=8|A|^{2}\frac{e^{-2a}}{\left( 1-e^{-2a}\right) ^{2}}%
\left( 3e^{-a}-e^{-3a}-2\right) \\
+4(AB^{\ast }+A^{\ast }B)e^{-a}-4|B|^{2}, \\
L_{\mathrm{defect}}=v|B|^{2}+\frac{w}{2}|B|^{4}~.
\end{gather}%
The substitution of the harmonic time dependence, $A(t)=Ae^{-i\omega
t},~B(t)=Be^{-i\omega t},$ where $A$ and $B$ are real amplitudes, casts the
Lagrangian into the following stationary form, with $C\equiv e^{-a}$:
\begin{gather}
L_{\mathrm{kin}}=4\omega \frac{A^{2}C^{2}}{\left( 1-C^{2}\right) ^{2}}%
+\omega B^{2}, \\
L_{\mathrm{self}}=2\sigma A^{4}\frac{C^{4}}{\left( 1-C^{4}\right) ^{2}}+%
\frac{\sigma }{2}B^{4}, \\
L_{\mathrm{coupl}}=8A^{2}\frac{C^{2}}{\left( 1-C^{2}\right) ^{2}}\left(
3C-C^{3}-2\right) \\
+8ABC-4B^{2}, \\
L_{\mathrm{defect}}=vB^{2}+\frac{w}{2}B^{4}~.
\end{gather}%
Fixing frequency $\omega $ and numerically solving the ensuing system of the
Euler-Lagrange equations, which follows from the effective Lagrangian, $%
\partial L/\partial A=\partial L/\partial B=\partial L/\partial C=0,$ one
can find the set of variational parameters $(A,B,a)$ and the norm
corresponding to ansatz (\ref{eq:ansatz}),
\begin{equation}
N=B^{2}\delta _{m,0}\delta _{n,0}+A^{2}\sum_{m,n}(1-\delta _{m,0}\delta
_{n,0})C^{2(|m|+|n|)}~.  \label{norm}
\end{equation}

\subsection{The linear defect}

Curves $N(\omega )$ for the discrete solitons produced by the VA, along with
their counterparts, obtained by dint of the direct numerical solution based
on the continuation from the anti-continuum limit (see Appendix A), are
displayed in Fig. \ref{fig1}, for the pure linear defect.

%\begin{figure}[th]
%\epsfig{file=V_N(w)_Delta01_lagrang.eps,width=4cm} %
%\epsfig{file=V_N(w)_Delta1_lagrang.eps,width=4cm} \label{fig1_V}
%\caption{(Color online) Comparison of the dependence of the norm on
%the frequency for the fundamental discrete solitons, as predicted by
%the VA, with the use of Eq. (\protect\ref{norm}) (thin blue lines),
%and its counterpart produced by a direct numerical solution of the
%stationary version of Eq. (\protect\ref{eq:dnls_def}) (solid and dashed black
%lines), for three different strengths of the pure linear defect
%($w=0$): $v=-1;0;1$ (from top to bottom; recall that $v>0$ and $v<0$
%correspond to the attractive and repulsive defects, respectively).
%The width of the defect in Eq. (\protect \ref{defects}) is $\Delta
%_{v}=0.1$ in (a) and $\Delta _{v}=1$ in (b). The nonlinearity is
%attractive ($\protect\sigma =+1$).} \label{fig1}
%\end{figure}

\begin{figure}[th]
\epsfig{file=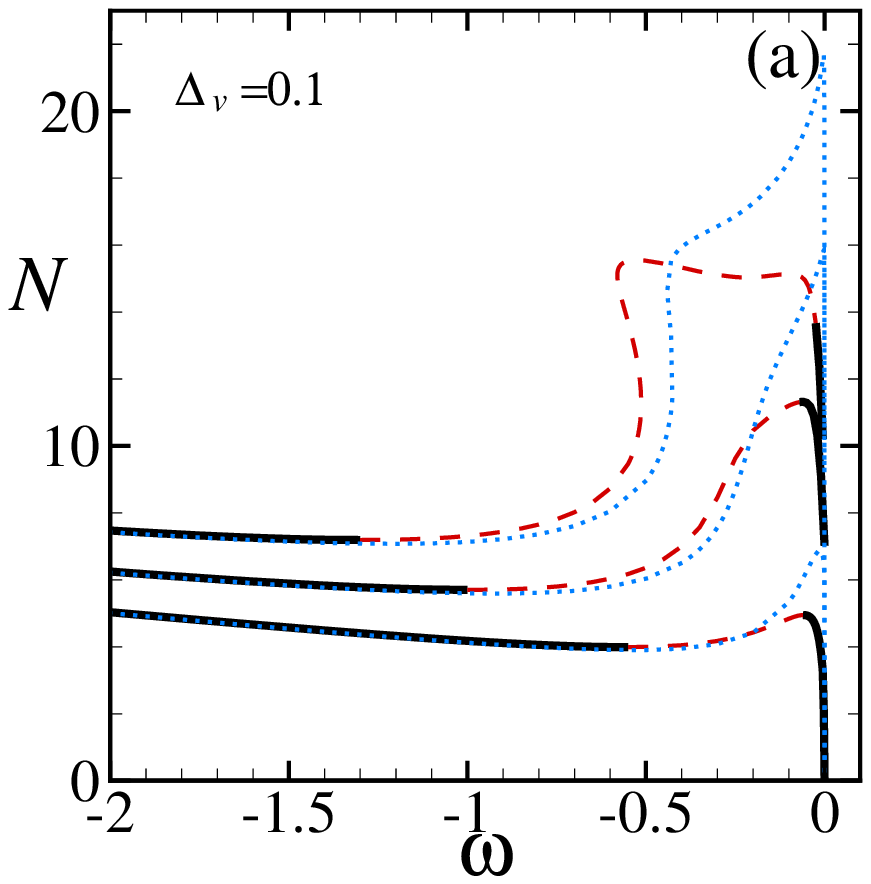,width=4cm} \epsfig{file=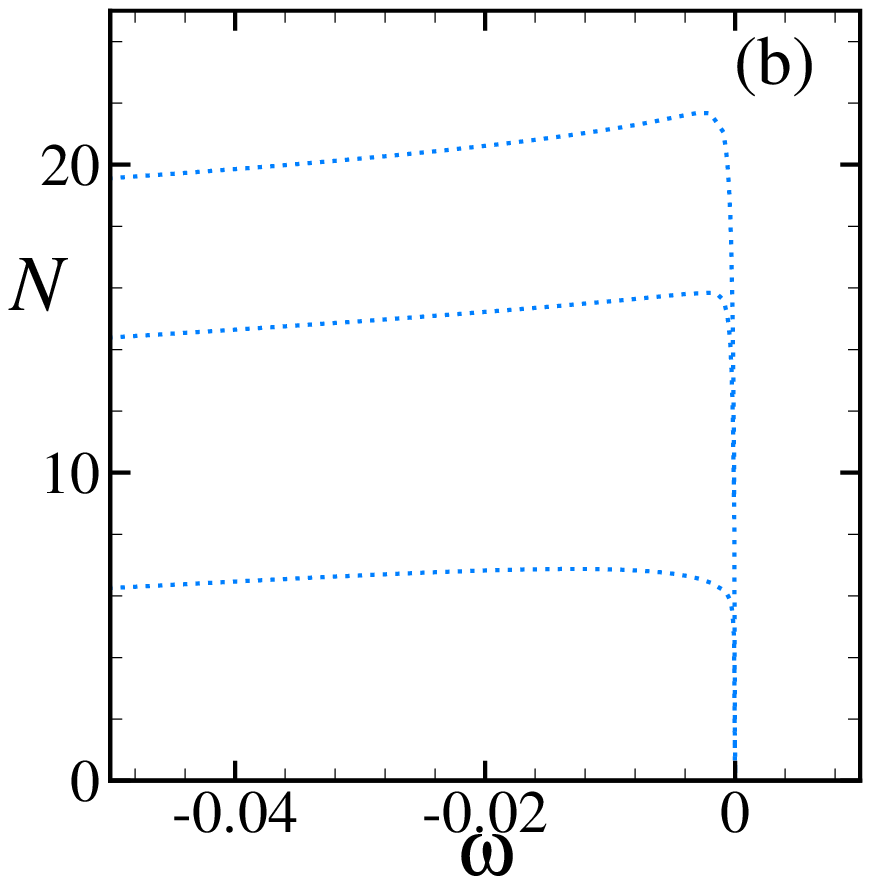,width=4cm}
\caption{(Color online) (a) Comparison of the dependence of the norm on the
frequency for the fundamental discrete solitons, as predicted by the VA,
with the use of Eq. (\protect\ref{norm}) (dotted blue lines), and its
counterpart produced by the numerical solution of the stationary version of
Eq. (\protect\ref{eq:dnls_def}) (solid black and dashed red lines), for
three different strengths of the linear defect ($w=0$): $v=-1;0;+1$ (from
top to bottom; recall that $v>0$ and $v<0$ correspond to the attractive and
repulsive defects, respectively). The numerical results were obtained for a
narrow defect, with width $\Delta _{v}=0.1$ in Eq. (\protect\ref{defects}).
The nonlinearity is attractive ($\protect\sigma =+1$). (b) Zoom of the plot
in (a), corresponding to the VA prediction for the norm given by Eq. (%
\protect\ref{norm}), in the vicinity of $\protect\omega =0$. In this and all
other figures, solid and dashed portions of the numerically generated curves
depict subfamilies of stable and unstable solitons, respectively.}
\label{fig1}
\end{figure}

Figure \ref{fig1}(a) shows that, in the case of $\Delta _{v}=0.1$, which is
practically tantamount to the defect localized at the single site, cf. Eq. (%
\ref{eq:dnls_def}), the VA-generated norm is in a good agreement with with
the numerical results far from the limit of $\omega \rightarrow 0$. At small
$\omega $, the soliton spreads out, and its actual shape deviates from the
exponential ansatz (\ref{eq:ansatz}), which leads to a discrepancy between
the variational and numerical results, as seen in Fig. \ref{fig1}(a). Also
due to restrictions on the size of the domain of calculation (in our case we
used $41\times 41$ grid points) the numerical curves cannot be continued to
extremely small values of $\omega $, therefore they are not presented in
Fig. \ref{fig1}(b).

\begin{figure}[th]
\epsfig{file=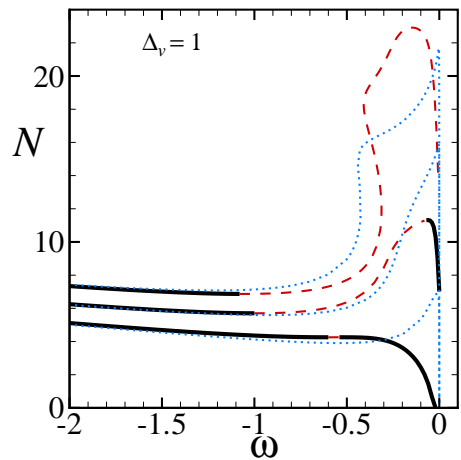,width=6cm}
\caption{(Color online) The same as in Fig. \protect\ref{fig1}(a), but with
the norm found numerically for the broad defect with width $\Delta _{v}=1$.}
\label{fig1_1}
\end{figure}
Naturally, the increase of the width of the defect to $\Delta _{v}=1$, which
makes the shape of defect (\ref{defects}) essentially different from that in
Eq. (\ref{eq:dnls_def}), leads to a stronger discrepancy in the limit $%
\omega \rightarrow 0$, as seen in Fig. \ref{fig1_1}.

We have checked the linear stability of the solutions along their existence
curves (see details in Appendix B). The results are shown by means of solid
and dashed portions in the figures displaying the respective numerically
generated $N(\omega )$ curves. Note that, here and in similar plots
displayed below, the soliton families may contain two distinct stability
segments separated by an instability region. It should be noted found that,
far from the limit of $\omega =0$, the Vakhitov-Kolokolov (VK) criterion, $%
dN/d\omega <0$ \cite{Panos}, correctly predicts the transition from the
stability to instability at points where $dN/d\omega $ changes its sign from
negative to positive. However, at small $\omega $, the formally applied VK
criterion only partially complies with the linear-stability results, which
may be an effect of boundary conditions on properties of very broad modes
corresponding to small $\omega $.

It is worthy to note that, as one can conclude from the comparison of Figs. %
\ref{fig1}(a) and \ref{fig1_1}, the increase of the width of the attractive
linear defect leads to a significant reduction of the instability region.
This trend is also corroborated by the VK criterion.

\subsection{The nonlinear defect}

\begin{figure}[th]
\epsfig{file=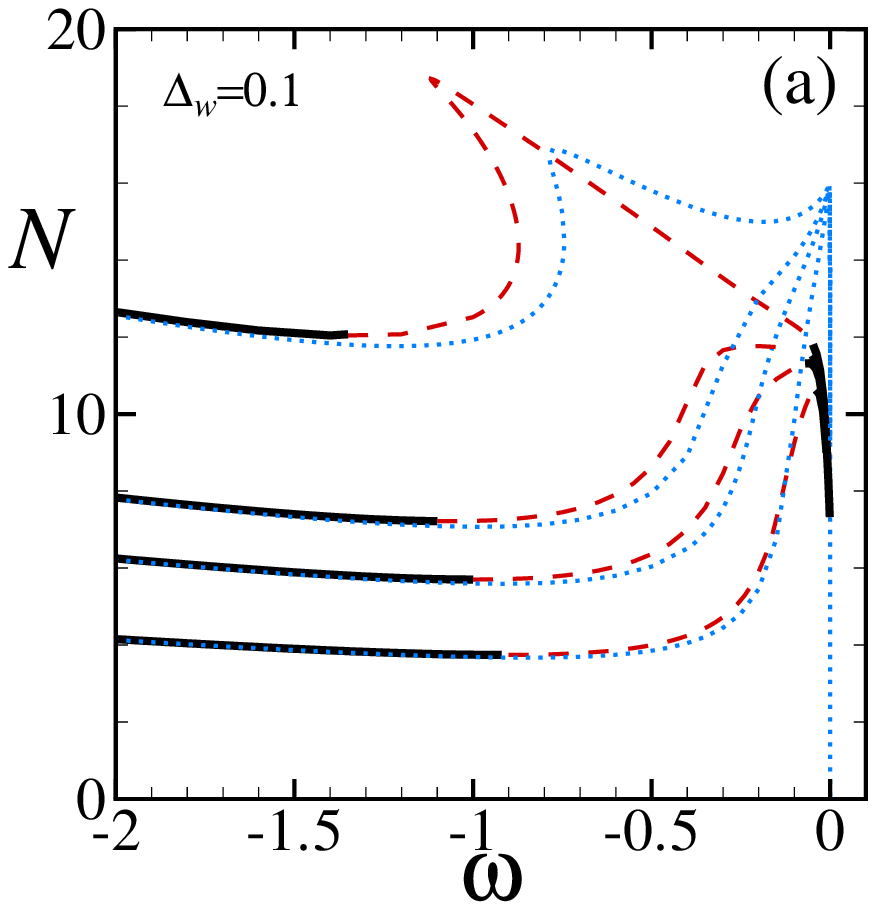,width=4cm} \epsfig{file=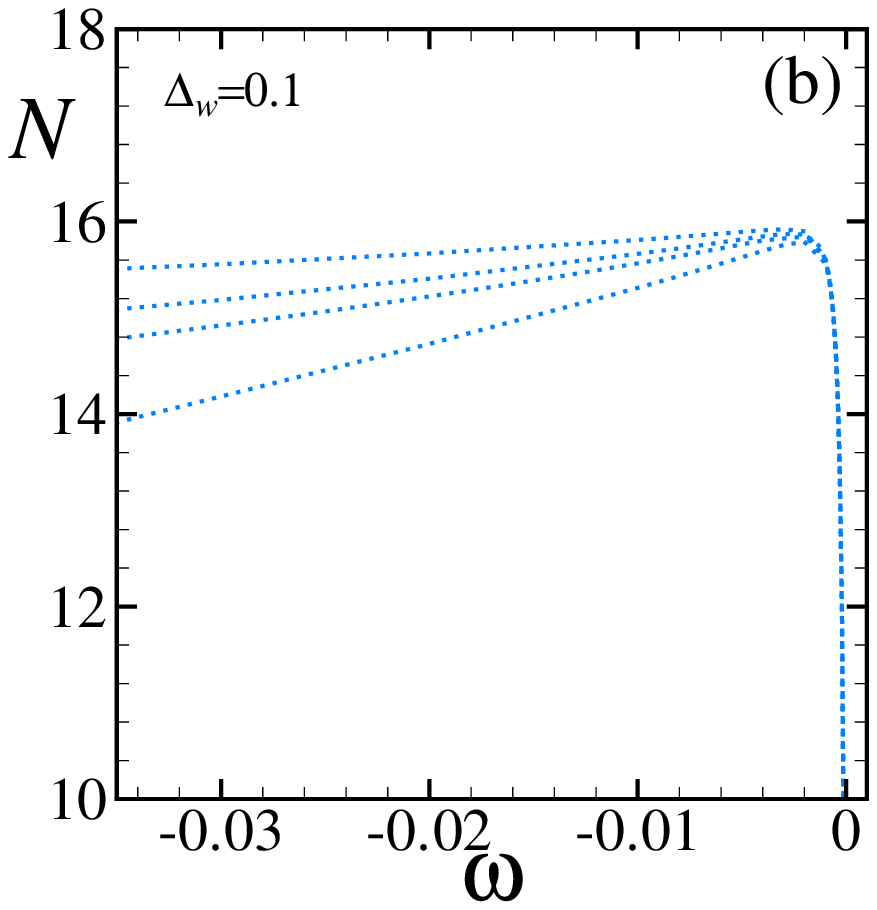,width=4cm}
\caption{(Color online) The same as in Fig. \protect\ref{fig1}, but for the
pure nonlinear defect ($v=0$), at different values of its strength: $%
w=-0.5;-0.2;0;+0.5$ (from top to bottom). The width of the defect
corresponding to the numerical curves is $\Delta _{w}=0.1$.}
\label{fig2}
\end{figure}

\begin{figure}[th]
\epsfig{file=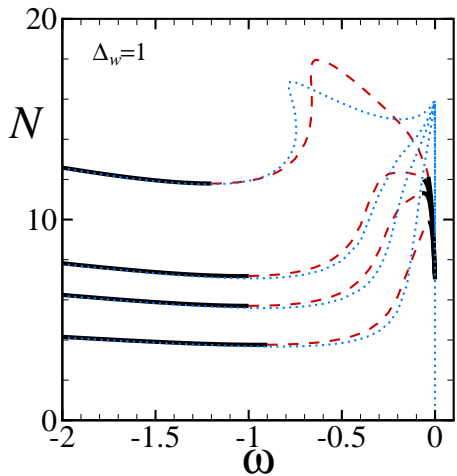,width=6cm}
\caption{(Color online) The same as in Fig. \protect\ref{fig2}(a), but for
the defect's width $\Delta _{w}=1$.}
\label{fig2_1}
\end{figure}

\begin{figure}[th]
\epsfig{file=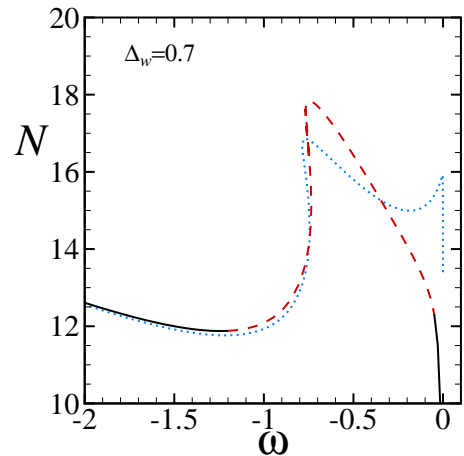,width=6cm}
\caption{(Color online) The comparison between the variational and numerical
results for the pure nonlinear defect with amplitude $w=-0.5$ and width $%
\Delta _{w}=0.7$. In this case, the overall discrepancy between the
VA-predicted and numerically found curves [which are defined as in Fig.
\protect\ref{fig2}(a)] attains its minimum.}
\label{fig2_07}
\end{figure}

Similar results were obtained for the pure nonlinear defect, i.e., with $v=0$
in Eqs. (\ref{defects}) and (\ref{eq:dnls_def}), as shown in Figs. \ref{fig2}
and \ref{fig2_1}. In particular, analyzing the results for the linear and
nonlinear defects with different widths, we have found that there is a
particular intermediate value of the width in the interval of $\Delta
_{v,w}\subset \left( 0.1,1\right) $, at which the discrepancy between the
numerical and variational solutions attains a minimum (in particular, it is $%
\Delta _{w}\approx 0.7$ for the nonlinear defect, see Fig.\ref{fig2_07}).
%\begin{figure}[th]
%\epsfig{file=W_N(w)_Delta01_lagrang.eps,width=4cm} %
%\epsfig{file=W_N(w)_Delta1_lagrang.eps,width=4cm} \label{fig1_W}
%\caption{(Color online) The same as in Fig. \protect\ref{fig1}, but
%for the pure nonlinear defect ($v=0$), at three values of its
%strength: $w=-0.5;0;0.5 $ (from top to bottom). The width of the
%defect is $\Delta _{w}=0.1$ in (a) and $\Delta _{w}=1$ in (b). }
%\label{fig2}
%\end{figure}

\begin{figure}[th]
\epsfig{file=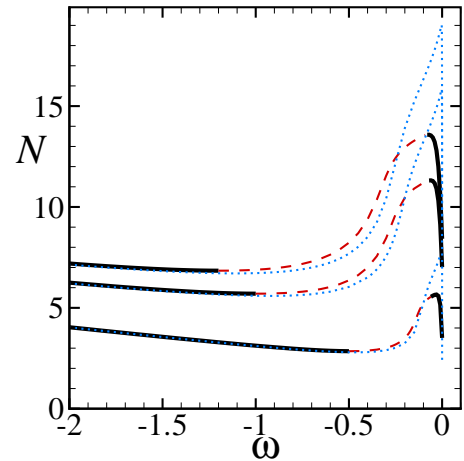,width=6cm}
\caption{(Color online) The same as in Figs. \protect\ref{fig1}(a) and
\protect\ref{fig2}(a), but for the fundamental discrete solitons in the 2D
lattice without defects ($v=w=0$), at three different values of the coupling
constant in Eq. (\protect\ref{timeDNLS}): $J=1.2;~1;~0.5$ (from top to
bottom).}
\label{fig3}
\end{figure}

For the sake of comparison, the VA-predicted and numerical curves $N(\omega
) $ are also shown in Fig. \ref{fig3} for the discrete solitons in the
defect-free 2D lattice (of course, these results are not different from
those reported in earlier works which were dealing with the 2D DNLS equation
without defects \cite{Weinstein,Panos}). Comparing these to Figs. \ref{fig1}
and \ref{fig2}, we conclude that the discrepancy between the VA and
numerical findings is actually \emph{smaller} in the presence of the linear
or nonlinear defect.

Finally, it is also interesting to consider the case of the \textit{linear
lattice} ($\sigma =0$), with all the nonlinearity concentrated only in the
form of a narrow defect (\ref{defects}) with $v=0$ and $w>0$. In Fig. \ref%
{fig_W0} , the corresponding dependence $N(\omega )$ is shown for $\Delta
_{w}=0.1$, which implies that the nonlinearity is actually concentrated at
the single site, $n=m=0$.

\begin{figure}[th]
\epsfig{file=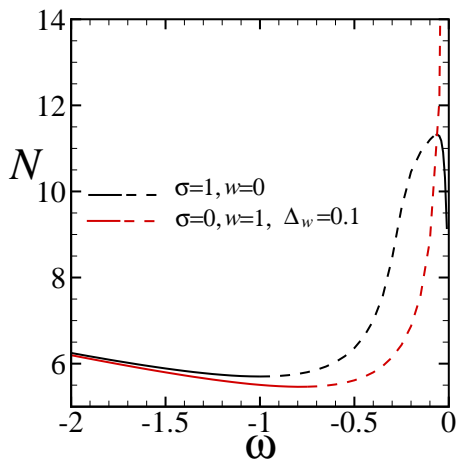,width=6cm}
\caption{(Color online) The comparison of the numerically found dependence $%
N(\protect\omega )$ in the \emph{linear} lattice ($\protect\sigma =0$), with
the nonlinearity represented \emph{solely} by the narrow nonlinear defect ($%
v=0,\Delta _{w}=0.1$)--red lines--with its counterpart in the usual
defect-free DNLS lattice (black solid and dashed lines).}
\label{fig_W0}
\end{figure}

\section{Localization-delocalization transition controlled by temporal
modulations of the linear and nonlinear defect}

\subsection{The variation of the linear defect}

In this section, we consider effects of the \textit{hysteresis} for the
discrete solitons, induced by the adiabatic variation of the strength of the
linear defect, $v(t)$, which eventually returns to the initial value, $%
v_{i}\equiv v(t=0)$, cf. Ref. \cite{BBK}:
\begin{equation}
v(t)=v_{f}+(v_{i}-v_{f})|1-2t/t_{f}|.  \label{Vt}
\end{equation}%
Here $v_{f}\equiv v(t=t_{f}/2)$ corresponds to the turning point, $t=t_{f}/2$%
, while the return time is $t=t_{f}$. Note that the total norm remains a
conserved quantity in the DNLS equation with the time-dependent strength of
the local defect.

\begin{figure}[th]
\epsfig{file=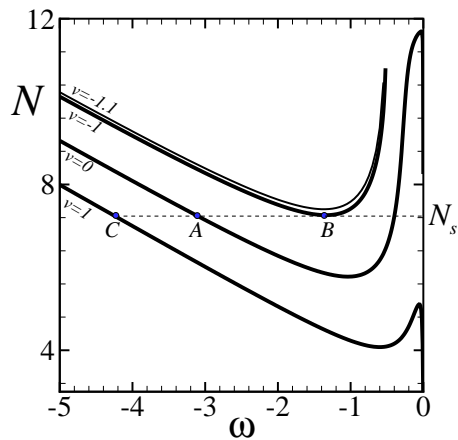,width=6cm}
\caption{(Color online) The illustration of the transformation of the
discrete soliton under the action of the adiabatic variation of the strength
of the linear defect.}
\label{fig4}
\end{figure}

\begin{figure}[th]
\epsfig{file=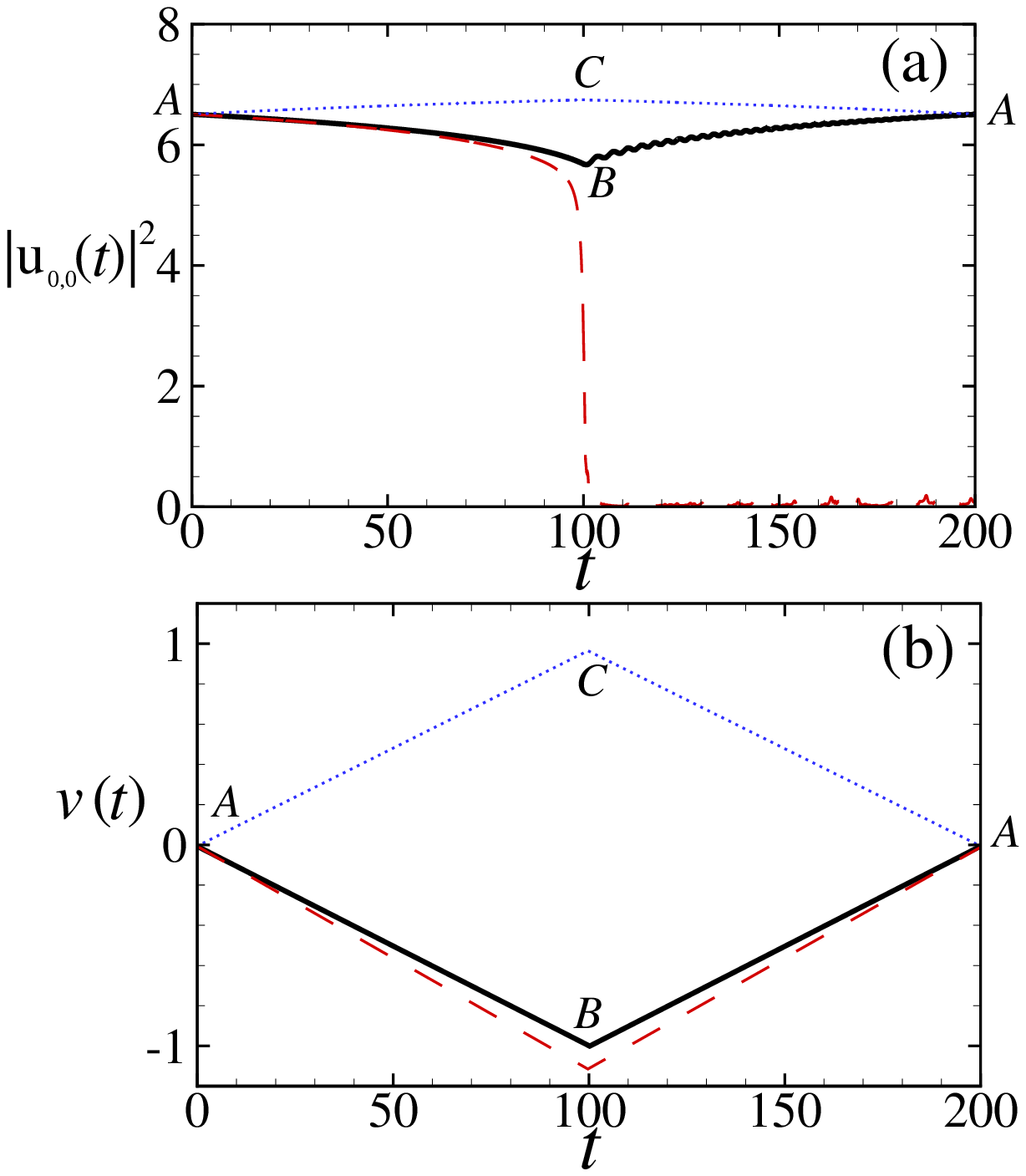,width=5cm}\epsfig{file=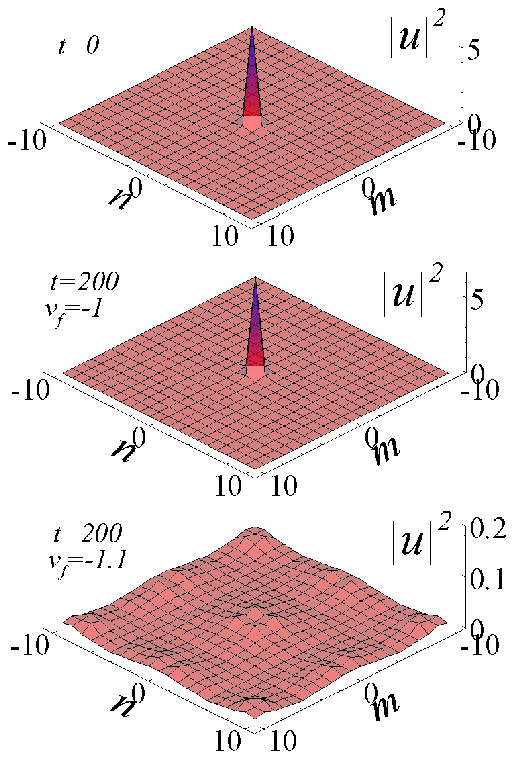,width=3.5cm}
\caption{(Color online) (a) The evolution of the density at the origin, $%
|u_{0,0}(t)|^{2}$, corresponding to the time-modulation scenarios displayed
in Fig. \protect\ref{fig4}. The dark solid and blue dashed-dotted lines
stand for the scenarios A-B-A (with $v_{f}=-1$) and A-C-A (with $v_{f}=1$),
which are shown in panel (b) and in Fig. \protect\ref{fig4}. The red dashed
line corresponds to the case when the strength $v(t)$ temporarily falls to a
value below the critical value, $v_{f}=-1.1<v_{\mathrm{cr}},$ which leads to
the delocalization. In the right panels, initial (top) and final density
profiles with $v_{f}=-1$ (middle) and $v_{f}=-1.1$ (bottom) are displayed.}
\label{fig5}
\end{figure}

First, we consider the soliton for $v_{i}=0$, with the norm, $N_{s}$ (point
A in Fig. \ref{fig4}), which precisely coincides with the minimum of the
existence curve, but for $v=-1$ (point B). This means that, by taking $%
v_{i}=0$ and $v_{f}=-1$ in Eq. (\ref{Vt}) at $t=t_{f}/2$, we may transfer
the soliton from point A to B. As long as the soliton stays on the existence
curve between these two points, it remains localized, featuring some changes
of the profile, see details of dynamics in Fig. \ref{fig5} (this is also
valid for the dynamics between points A and C). However, if, in the
framework of the same scenario, the amplitude of the defect falls below the
critical value, $v_{f}<v_{\mathrm{cr}}$, which in the present case is $v_{%
\mathrm{cr}}=-1$, then the minimum of the existence curve turns out to be
higher than the norm of the evolving mode, which is expected to trigger the
delocalization.

These assumptions were verified by means of the direct integration of Eq. (%
\ref{timeDNLS}) with the strength of the linear defect $v$ varying in time
according to Eq. (\ref{Vt}). The result is presented in Fig. \ref{fig5},
where the evolution of the density at the origin, $|u_{0,0}(t)|^{2}$, is
shown for three values of the $v_{f}$, along with the corresponding initial
and final profiles.

\subsection{The variation of the nonlinear defect}

Here we aim to consider effects produced by the adiabatic variation of the
strength of the nonlinear defect, following the same scenario as in Eq. (\ref%
{Vt}):
\begin{equation}
w(t)=w_{f}+(w_{i}-w_{f})|1-2t/t_{f}|,  \label{wt}
\end{equation}%
with $w_{i}=w(t=0)$ and $w_{f}=w(t=t_{f}/2)$ corresponding to the initial
strength and its value at the turning point, $t=t_{f}/2$, with the return to
the initial value at $t=t_{f}$. As in the previous case, the norm of the
solution remains conserved in the course of the evolution.

\begin{figure}[th]
\epsfig{file=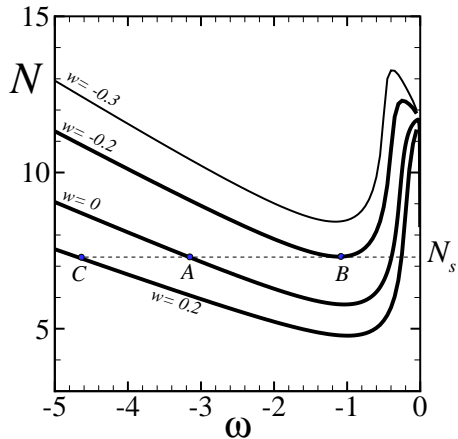,width=6cm}
\caption{ (Color online) The illustration of the soliton transformation
under the action of the adiabatic variation of the strength of the nonlinear
defect.}
\label{fig6}
\end{figure}

\begin{figure}[th]
\epsfig{file=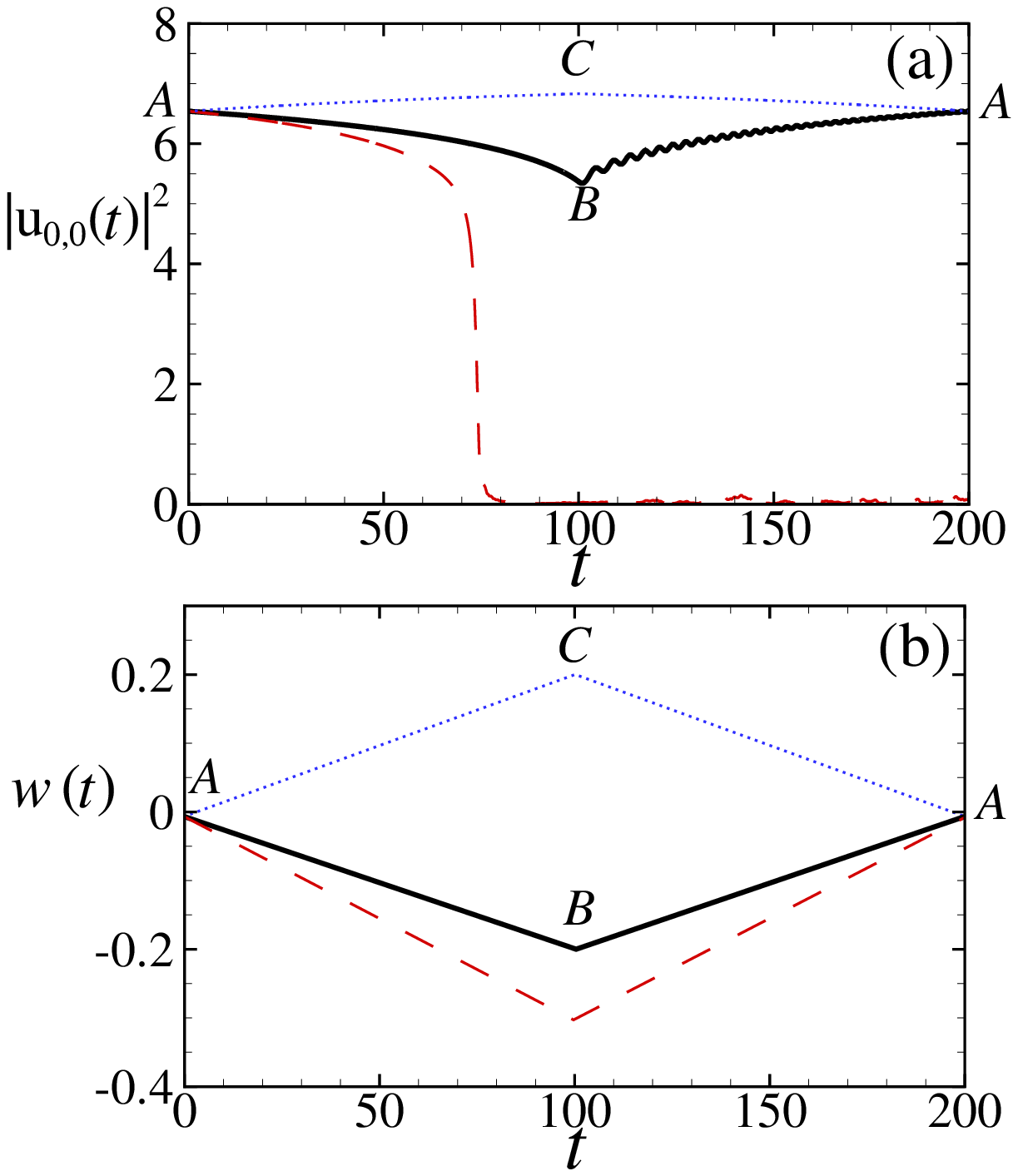,width=5cm}\epsfig{file=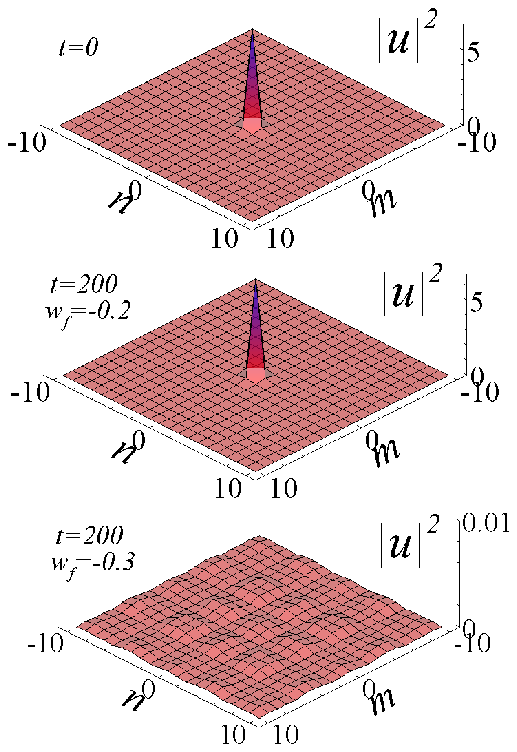,width=3.5cm}
\caption{ (Color online) The same as in Fig. \protect\ref{fig5}, but for the
temporal modulation of the nonlinear defect.}
\label{fig7}
\end{figure}

Again we start with the solution at $w_{i}=0$ (point A in Fig. \ref{fig6})
with the norm which coincides with the minimum of the existence curve for $%
w=-0.2$ (point B). This means that, by taking $w_{i}=0$ and $w_{f}=-0.2$ in
Eq. (\ref{wt}) at time $t=t_{f}/2$, we will transfer the soliton from point
A to point B, see Fig. \ref{fig6}. If the soliton stays on the existence
curve between these two points, the solution always remains localized (see
the dynamical picture in Fig. \ref{fig7}). However, if, for the same initial
condition, the strength of the nonlinear defect is allowed to drop below the
critical value, $w_{f}<w_{\mathrm{cr}}$, which in the present case is $w_{%
\mathrm{cr}}=-0.2$, then the minimum of the existence curve becomes higher
than the norm of the solution, which again is expected to trigger the
transition to the delocalization.

These assumption were verified through the direct integration of Eq. (\ref%
{timeDNLS}) with the strength of the nonlinear defect, $w$, varying in time
according to Eq. (\ref{wt}). The results are presented in Fig. \ref{fig7},
where the evolution of the density at the origin, $|u_{0,0}(t)|^{2}$, is
shown for three values of $w_{f}$.

\section{Conclusion}

In this work, we have considered the static and dynamical properties of 2D
discrete solitons in the nonlinear lattice described by the DNLS (discrete
nonlinear Schr\"{o}dinger) equation, which includes the local linear or
nonlinear defect. The solitons trapped around the defects were investigated
using both the VA (variational approximation) and numerical methods, the VA
showing good agreement with the numerical findings for sufficiently narrow
solitons. Then, the analysis was extended to the model with the linear and
nonlinear defects whose strength was subject to the slow variation in time.
In the latter case, one of the possibilities is the controlled onset of the
transition to delocalization.

The analysis reported in this work can be extended by considering
combinations of linear and nonlinear defects, searching for
symmetric, antisymmetric, and asymmetric modes trapped by symmetric
pairs of defects, and analyzing other ``management" schemes, with
the defect strength subject to a time-periodic modulation.

\section*{Acknowledgments}

V.A.B. acknowledges support from the FCT grant, PTDC/FIS/64647/2006. B.A.M.
appreciates hospitality of Centro de F\'{\i}sica do Porto (Porto, Portugal).

\section*{Appendix A}

Here we discuss some details on the continuation of solutions from the
anti-continuum (AC)\ limit, which corresponds to the uncoupled lattice
described by Eq. (\ref{timeDNLS}) and (\ref{eq:dnls_def}) with $J=0$ \cite%
{Panos}. Looking for stationary modes in the form of Eq. (\ref{omega}), in
the AC limit one obtains obvious exact solution for the one-site fundamental
discrete soliton,
\begin{equation}
U_{m,n}=\left\{
\begin{array}{ll}
\pm \mathcal{A}, & \mbox{if}\quad m=n=0; \\
0, & \mbox{otherwise}.%
\end{array}%
\right.
\end{equation}%
where $\mathcal{A}\equiv \sqrt{-\left( \omega +v\right) /\left( \sigma
+w\right) }$, see Ref. \cite{Alfimov} and references therein. Then, by using
the standard Newton-Raphson method, we gradually increase the coupling
constant from $J=0$ to $J=1$, also increasing the width of the defect if
needed [to proceed from the narrow defect (\ref{eq:dnls_def}) to the broader
one (\ref{defects})]. After that, by changing frequency $\omega $ of the
solution, we obtain the dependence of norm $N$ on $\omega $, and also
analyze the linear stability of the corresponding solutions, see the next
Appendix.

\section*{Appendix B}

To analyze the stability of solitons, we consider perturbed solutions,
\begin{equation}
u_{m,n}(t)=\left[ U_{m,n}+a_{m,n}(t)\right] e^{-i\omega t},
\end{equation}%
Representing the small perturbations as $a_{m,n}(t)=\left( \alpha
_{m,n}+i\beta _{m,n}\right) e^{\lambda t}$, substituting this into Eq. (\ref%
{timeDNLS}) and separating real and imaginary parts, we derive the following
system,
\begin{equation}
\lambda \left(
\begin{array}{l}
\alpha  \\
\beta
\end{array}%
\right) =\left(
\begin{array}{ll}
0 & L_{1} \\
-L_{3} & 0%
\end{array}%
\right) \left(
\begin{array}{l}
\alpha  \\
\beta
\end{array}%
\right)   \label{sys}
\end{equation}%
where $\lambda $ is the stability eigenvalue, and operators $L_{1}$ and $%
L_{3}$ are
\begin{eqnarray}
L_{1} &=&\Delta _{2}+V_{m,n}-\omega +(\sigma +W_{m,n})\left( U_{m,n}\right)
^{2},  \label{L1} \\
L_{3} &=&\Delta _{2}+V_{m,n}-\omega +3(\sigma +W_{m,n})\left( U_{m,n}\right)
^{2}.  \label{L3}
\end{eqnarray}%
Multiplying the first equation in system (\ref{sys}) by $-\lambda $ and than
using the second equation, one arrives at the eigenvalue problem,
\begin{equation}
\Lambda \alpha =L_{1}L_{3}\alpha ,
\end{equation}%
where $\Lambda =-\lambda ^{2}$. The underlying stationary solution, $U_{m,n}$%
, is unstable if the spectrum of $\Lambda $ includes complex or negative
real values.

\end{document}